\documentclass[apj]{emulateapj}

\def\sun{\hbox{$\odot$}}

\slugcomment{{\sc Accepted to ApJ:} September 19, 2008}

\shorttitle{Spectro-interferometry of Mira Variables from 1.1 to 3.8 microns}
\shortauthors{Woodruff et al.}

\begin{document}

\title{The Keck Aperture Masking Experiment: spectro-interferometry of 3 Mira Variables from 1.1 to 3.8\,$\mu$m }

\author{H.C.~Woodruff\altaffilmark{1}, 
	M. J.~Ireland\altaffilmark{1},
	P. G.~Tuthill\altaffilmark{1},
	 J. D.~ Monnier\altaffilmark{2}, 
	T. R.~Bedding\altaffilmark{1}, 
        W. C.~Danchi\altaffilmark{3},
        	M.~Scholz\altaffilmark{1,4}, 
	C. H.~Townes\altaffilmark{5}, and 
	P. R. ~Wood\altaffilmark{6}}

\altaffiltext{1}{Sydney Institute for Astronomy (SIFA), School of Physics, University of Sydney, NSW 2006, Australia}
\altaffiltext{2}{University of Michigan at Ann Arbor, Department of Astronomy, 
                 500 Church Street, Ann Arbor, MI 48109-1090, USA}
\altaffiltext{3}{NASA Goddard Space Flight Center, Exoplanets and Stellar Astrophysics, 
                 Code 667, Greenbelt, MD 20771, USA }
 \altaffiltext{4}{Zentrum f\"ur Astronomie der Universit\"at Heidelberg (ZAH), Institut f\"ur Theoretische Astrophysik, Albert-Ueberle-Str. 2, 69120 Heidelberg, Germany}
  \altaffiltext{5}{University of California at Berkeley, Space Science Laboratory, Berkeley, CA 94725-7450, USA}
  \altaffiltext{6}{Research School of Astronomy and Astrophysics, Australian National University, Cotter Road, Weston Creek ACT 2611, Australia}

\begin{abstract}
We present results from a spectro-interferometric study of the Miras $o$ Cet, R Leo and W Hya  
obtained with the Keck Aperture Masking Experiment from 1998 Sep to 2002 Jul.
The spectrally dispersed visibility data permit fitting with circularly symmetric brightness profiles such as a simple uniform disk.
The stellar angular diameter obtained over up to $\sim$\,450 spectral channels spaning the region 1.1-3.8\,$\mu$m is presented.
Use of a simple uniform disk brightness model facilitates comparison between epochs and with existing data and theoretical models.
Strong size variations with wavelength were recorded for all stars, probing zones of H$_2$O, CO, OH, and dust formation.
Comparison with contemporaneous spectra extracted from our data show a strong anti-correlation between the observed angular diameter and flux.
These variations consolidate the notion of a complex stellar atmosphere consisting of molecular shells with time-dependent densities and temperatures.
Our findings are compared with existing data and pulsation models. 
The models were found to reproduce the functional form of the wavelength vs. angular diameter curve well,
although some departures are noted in the 2.8-3.5\,$\mu$m range.

\end{abstract}


\section{Introduction}

Due to their large diameters and high luminosities, Mira variables are a favourite target for
optical and near-infrared (NIR) interferometric observations, which have shown
dependencies of diameter on wavelength, pulsation phase and pulsation cycle (e.g., \citealt{HAN95}; \citealt{VANB}; \citealt{PER}; \citealt{JAC00}; \citealt{YOU00}; \citealt{HOF02}; \citealt{THOM}; \citealt{IRE2004}; \citealt{WOO};\citealt{WOO08}).
This wealth of interferometric information has advanced the studies of molecular 
and dust abundances in the atmosphere (e.g. \citealt{ISTW,ISW,IS06}),
the pulsation mode of these stars (e.g., \citealt{WOO}; \citealt{FED05}), 
and photospheric/circumstellar asymmetries (e.g.  \citealt{RAG06}) as well as
the characteristics of the circumstellar environment (e.g. \citealt{DAN94}).\\

Miras  are long period variables (LPVs) of roughly one solar mass which evolve along the Asymptotic Giant Branch (AGB)
and are characterized by well-defined pulsation periods and large pulsation amplitudes (up to $\Delta V\approx9$).
The extended atmospheres of these late-type giants show phase-dependent density and temperature stratifications
which, together with complicated intensity profiles and the presence of molecular layers, leave no simple observable quantity
to define the structure of the star's envelope.
The pulsations also help drive strong winds thus enriching the interstellar medium
with molecules such as H$_2$O, CO, TiO and SiO (\citealt{TSU97}; \citealt{TLS}; \citealt{OHN04}).
A better understanding of their atmospheric structures will thus further the understanding of these late stages of stellar evolution,
and might help shed light on one of the most important sources for the chemical enrichment of the interstellar medium.

To fully characterize the structure of a Mira star's atmosphere, one would need the complete intensity profile 
at all wavelengths and all pulsation phases, recorded over a large number of pulsation cycles. 
The last decade has shown a multitude of angular diameter measurements of Miras in various wavelengths,
with some instruments returning impressive resolution of the brightness profile (e.g. IOTA: \citealt{RAG06}, VLTI: \citealt{WIT08}, etc).
The difficulty lies in gathering multi-phase, multi-wavelength data on a single object.
Here, we present multi-epoch, spectro-interferometric observations of the nearby Mira stars $o$~Cet, R~Leo and W~Hya covering
the NIR spectrum from 1.1-3.8\,$\mu$m near-simultaneously, with one-dimensional spatial information out to the diffraction limit of the Keck~I telescope.
With our data set we are able to sample phase-dependent molecular stratification and help probe the structure
of the stellar atmospheres.\\

\section{Observations and Data Reduction}\label{observations}
\subsection{Aperture-masking Observations}

Our sample of 3 Miras (W~Hya is sometimes classified as a semi-regular pulsator with strong Mira characteristics) were chosen for their large angular diameters and NIR brightness (see Table \ref{objects}).
Observations were performed with the 10\,m Keck I telescope using the Near Infrared Camera (NIRC).
The telescope pupil was converted into a sparse interferometric array by placing an aperture mask in the beam 
in front of the infrared secondary mirror, allowing
the recovery of the Fourier amplitudes and closure phases for baselines up to 10\,m.

\begin{deluxetable}{lccc}
\tablewidth{0pt}
\tablecaption{\label{objects}
Observed objects
}
\tablehead{
\colhead{Name} & \colhead{Period} & \colhead{Spectral} &  \colhead{Distance}\\
			&\colhead{[days]}	&\colhead{Type Range}		& \colhead{[pc]}\\
}
\startdata
W~Hya	&      385  & M7.5-9ep & $104\pm12$ \\
R~Leo	&	312	& M6-9.5e & $111\pm17$ \\
$o$~Cet 	&	332	& M5-9e & $92\pm10$ \\
\enddata
\tablerefs{Object Period from the American Association of Variable Star Observers (AAVSO) visual light curves (A.A. Henden et al. 2006, private communication), 
M Spectral Type Range from \cite{SLOANPRICE} and distances from \cite{LEE07}
}
\end{deluxetable}

\begin{deluxetable*}{lcccc}
\tablewidth{0pt}
\tablecaption {\label{grisms}  
Table of grisms}
\tablehead{
 \colhead{Grism} &  \colhead{Keck/NIRC blocking} & \colhead{Grism wavelengths} &   \colhead{effective wavelengths } & \colhead{spectral resolution} \\ 
 				&\colhead{Filter Name}	&	\colhead{[$\mu$m]}	&\colhead{[$\mu$m]}                  &\colhead{[$10^{-3}\,\mu$m]}\\        
}
\startdata
GR\,150	&JH		&    0.82--2.08   	&     1.0--1.6	  & $\sim5$\\ 
GR\,120 	&HK 		&    1.05--2.58		&    1.4--2.5  & $\sim6$  \\
GR\,60	&KL		&    1.98--4.75		&    2.1--4.5 & $\sim10$\\

\enddata

\end{deluxetable*}

\begin{deluxetable*}{lccccc}
\tablewidth{0pt}
\tablecaption{\label{tbl-obs}
Observations of Mira stars in our sample, including cycle and visual phase $\Phi$ (cycle=0 for first observation, visual phase=0 at maximum light)}
\tablehead{
\colhead{Object name} & \colhead{Date} & \colhead{JD}  &  \colhead{$\Phi$} &\colhead{Grisms}&\colhead{Slit width} \\
 			& &	 \colhead{$-$2450000}  &		&                 &\colhead{[pixels]} 	                           \\
 }
\startdata
$o$ Cet &1998Sep29 	&   1056 & 0.71	&GR60,120,150& 1.5, 1.5, 1.5\\
 & 2002Jul23 &   2479 & 4.98  &GR60,120& 1.5, 3.5\\
 R Leo &1999Feb04&1213&0.49&GR120,150& 3.5, 3.5\\
 & 1999Apr25&1295&0.75&GR60,120,150& 4.5, 4.5, 4.5\\
W Hya & 1999Feb05&1213&0.58&GR60,120,150& 2.5, 3.5, 3.5\\
 & 1999Apr25&1295&0.79&GR60,120,150&4.5, 4.5, 4.5\\
 & 2000Jan25&1570&1.53&GR60,120,150&1.5, 3.5, 4.5\\
\enddata
\end{deluxetable*}

\begin{deluxetable}{lccc}
\tablewidth{0pt}
\tablecaption{\label{calibrators}
Reference stars with spectral types and estimated diameters
}
\tablehead{
\colhead{Reference star} & \colhead{Spectral} & \colhead{Adopted UD} & \colhead{Reference}\\
				&\colhead{Type} 	&\colhead{Angular Diameter [mas]} &\\
}
\startdata
2 Cen 		&	M4.5III	 & 	$14.7$		&1 \\
$\alpha$~Cet 	&	M1.5III	 & 	$11.7\pm0.6$	&2 \\
$\alpha$~Hya 	&	K3II-III	 & 	$9.1\pm0.1$	&3 \\
$\alpha$~Lyn 	&	K7III		 & 	$7.2\pm0.6$	&2\\
$\alpha$~Tau & 	K5III		&	$19.7\pm0.1$	&4\\
\enddata
\tablerefs{(1) \cite{DUM98}; (2) \cite{DYCK98}; (3) \cite{MOZ03}; (4) \cite{PER98}
}

\end{deluxetable}

Our experimental methods are derived from the highly 
successful aperture masking program at the Keck observatory which, in addition to studies of evolved pulsating stars and giants
\citep{TUT00b,MON04,WOO08}, has also delivered advances in stellar astrophysics
ranging from young stellar objects \citep{DAN01,TUT02}
to dusty mass-loss shrouds in proto-planetary nebulae and massive stars
\citep{MON00,TUT02b,TUT06}.
A full description of the experiment, including a discussion of the conceptual
principles and signal-to-noise considerations underlying masking interferometry,
is given in \citet{TUT00}, while further discussion of systematics and seeing
induced errors can be found in \citet{MON04}.

\begin{figure}[htbp]
\begin{center}

\epsscale{1.0}
\plotone{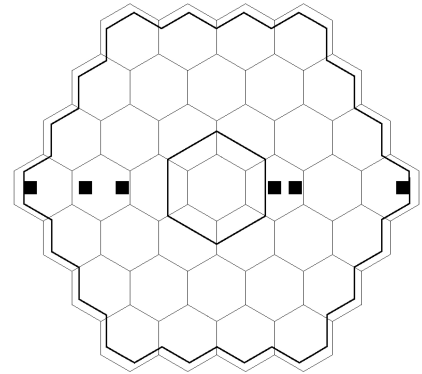}

\caption{Pupil geometry used for the Keck masking $+$ grism experiment.
A 6-hole non-redundant linear array (dark squares) is shown overlaying a scaled version of the segmented Keck primary mirror.
The boundary of the f/25 secondary mirror as projected on the primary is represented as the bold black line.}

\label{mask}
\end{center}

\end{figure}

In a significant extension to the capabilities of our previous experimental
setup, data presented here were obtained in a spectrally-dispersed manner 
delivering continuous wavelength coverage across the infrared J, H and K
bands (from 1.0\,$\mu$m to 3.7\,$\mu$m).
The use of a one-dimensional non-reduntant aperture mask, as depicted in 
Figure \ref{mask}, in combination with a cross-dispersing grism
element already installed in the NIRC camera \citep{MAT96}, resulted in
spectral information being encoded in one dimension on the readout array, 
while spatial interference fringes were recorded in the other dimension.
Note that with this setup, the full two-dimensional imaging data recovered
in the previous experiment was sacrificed for one-dimensional brightness
profiles across the waveband. 

Figure \ref{mask} shows the aperture mask, which was designed to fit within the
Keck's segmented primary. 
The design is a trade-off between uniform Fourier coverage and avoiding features 
within the pupil, such as the central obstruction and the boundaries between
mirror segments.
Although the mask only passes a small amount of the light gathered by the 10\,m 
primary mirror, and still more light is lost at the slit entrance, 
the LPV targets discussed in this paper are among the most luminous objects
in the near-infrared sky and the low overall throughput did not adversely 
affect the signal-to-noise of the final data.
On the contrary, we found that there was even sufficient flux recorded on the
shoulders and in between the traditional near-infrared band windows, where 
atmospheric absorption is high, so that nearly continuous wavelength coverage 
could be obtained, with the exception of the strong telluric emission features, due to water, between 2.6 and 2.9\,$\mu$m. 

The NIRC camera has three grisms intended to cover the J-H, H-K and
K-L spectral regions (see Table \ref{grisms}).
All three of these grisms were used and, as the reasonable degree
of wavelength overlap between them provided a useful check on the reliability 
of the data.
Also, four different slit widths (1.5, 2.5, 3.5, 4.5\,pixels as projected on the detector) 
were used in order to strike a balance between spectral resolution and available flux.
The grisms' wavelength resolution are given in Table \ref{grisms} and are to be multiplied
with the slit width used in the observation (see Table \ref{tbl-obs}) convolved with 1 pixel of the detector
to obtain the ``true'' wavelength resolution of the instrument.

\begin{figure*}[htbp]
\begin{center}
\epsscale{1.0}

\plottwo{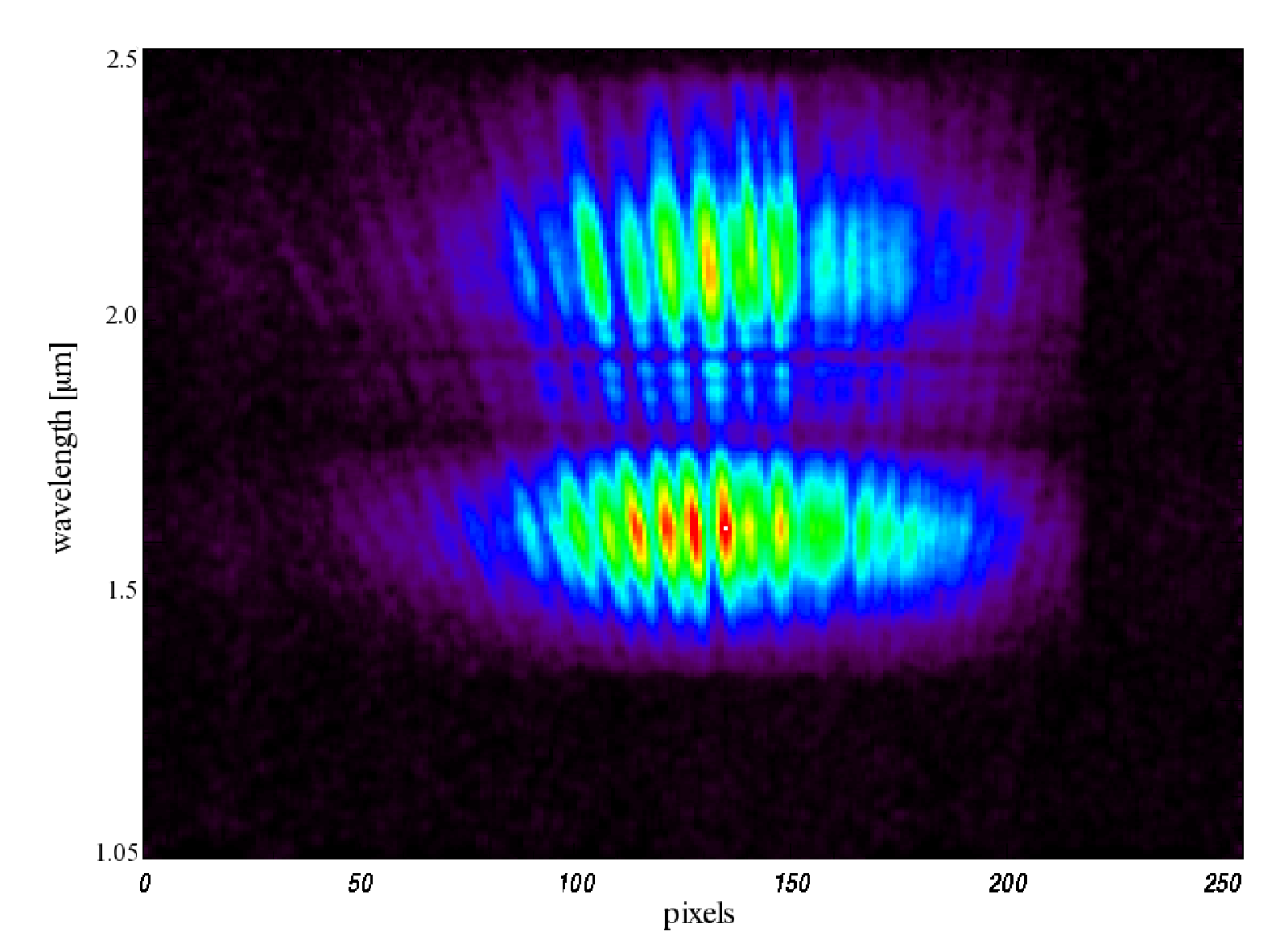} {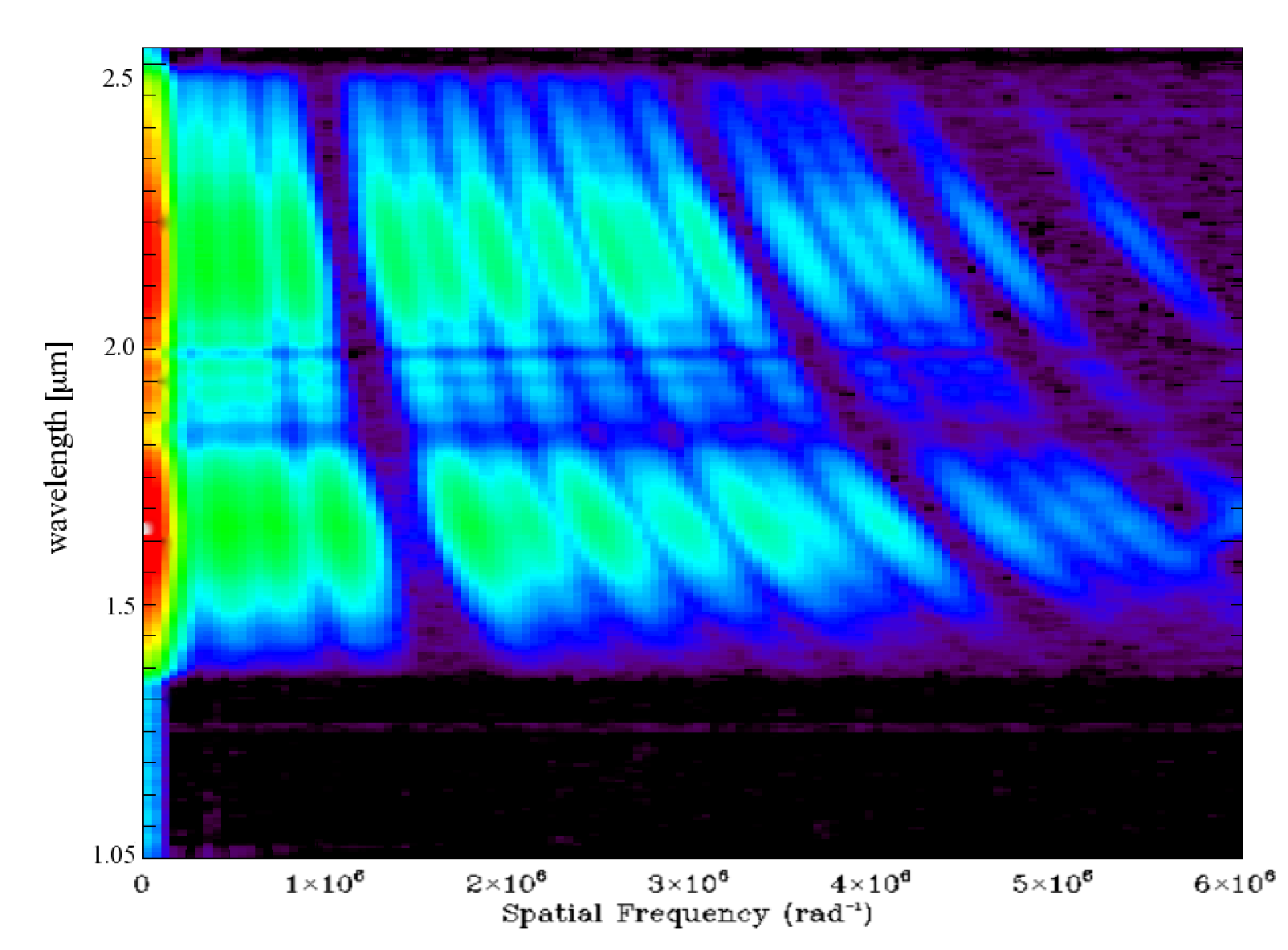}

\caption{Spectrally dispersed interferogram of W Hya taken with the GR120 grism ({\it left}). The interference pattern 
generated by the 6-hole linear mask (see Fig. \ref{mask}) along the abscissa is dispersed with wavelength along the ordinate.
The power spectrum ({\it right}) is the result of averaging the power spectra of 100 such data frames. Note the regions of attenuated
signal due telluric absorption in both the interferogram and the power spectrum.}

\label{speckle_ps}
\end{center}

\end{figure*}

After passing the mask and being dispersed in the NIRC spectrographic mode, 
the beams were focused on the $256\times256$ pixel array at a scale of 
20.57\,mas\,pixel$^{-1}$. This is sufficient to
Nyquist-sample data collected for $2\,\mu$m or longer wavelengths.
An example of a typical short-exposure ($T_{int}$=0.14\,s) frame is shown in
the left panel of Figure~\ref{speckle_ps}. 
Following the methods of \citet{TUT00}, 100 of these frames could be 
collected with reasonable efficiency (duty cycle $\sim$20\%) into a 
data cube for processing, to statistically calibrate the 
atmospherically degraded point-spread function. 

The power spectrum accumulated over a cube of 100 frames is illustrated
in the right panel of Figure~\ref{speckle_ps}.
Power on fifteen baselines can be clearly seen as peaks appearing at
discrete spatial frequencies that vary smoothly with wavelength. 
By recording the power on each baseline for a given wavelength bin, a 
one-dimensional visibility function could be recovered for that channel.
When calibrated for the total flux level in each channel, these raw 
1-dimenional $V^2$ data could be collated and fitted with any convenient
model (e.g. a uniform disk; model fitting is discussed in Section \ref{UDC}),
yielding a continuous sampling of the angular size of the target as a 
function of wavelength. 
Before such fitting could take place, calibration for the telescope-atmosphere
transfer function was needed. For a number of instrumental reasons, this 
procedure turned out to be more complicated than in the past, as described
below.

At wavelengths shorter than  $2\,\mu$m, the longest baseline corresponds to spatial frequencies
higher than the Nyquist sampling condition. This power is aliased back into the power 
spectrum at lower spatial frequencies.
This effect can be observed in the power spectrum in Fig. \ref{speckle_ps} as an apparent
reflection of the long-baseline power peaks when the wavelength is below $\approx2\,\mu$m.
Where this wrapped signal overlaps with power from shorter baselines, the data become
confused and were discarded. 

In common with established practice in interferometry, observations of PSF
calibrator stars with smaller and well-characterized angular diameters 
were interleaved with those of the science target (see Table \ref{calibrators}).
However, for observations with the grism setup utilized here, an 
observational difficulty compromised this ``standard'' calibration process.
Due to the experimental requirement for a stationary optical path between 
the mask and detector, the Keck image rotator was not used for any masking experiment.
This precluded use of the normal telescope guiding system, and
each series of 100 exposures (lasting of up to a couple of minutes) were taken with the telescope 
in a blind tracking mode. While small drifts in tracking had no impact on the 
original masking experiment \citep{TUT00}, this was not the case when the grism 
was employed because wander of the stellar image across the slit was
found to result in significant modification to the optical transfer function.
This motivated the construction of an alternate calibration strategy outlined in section \ref{UDC}.

The spectrally dispersed visibility data resolve only one spatial
dimension at a single position angle for each target and hence permits fitting with circularly symmetric brighness
 profiles such as a simple uniform disk.
The spectro-interferometric data were recorded at 6 different epochs spanning more than 4 years.
Table \ref{tbl-obs} lists these observations.

\subsection{Uniform Disk Diameters and Calibration}\label{UDC}

As discussed in \cite{WOO08}, the true stellar intensity profile is not a uniform disk (UD),
but shows a complex center-to-limb variation that varies with wavelength,
pulsation phase and cycle.
Fitting the observed 1-d visibilities with a simple profile like a UD
provides a useful estimate of the apparent  change of size as a function of wavelength.  
Since we are only resolving low-resolution structure in the star's intensity profile, it becomes difficult to
differentiate between a UD, a fully-darkened disk, or a Gaussian. Any one of these simple models
will show a similar relative change in diameter with wavelength, so
we chose UD diameters to allow comparison of findings with existing literature and to avoid
the difficulties encountered when fitting more sophisticated profiles
(cf. \citealt{HSW,SCH03}).

All spectrally-dispersed observations using the grism were accompanied on 
the same night with orthodox two-dimensional aperture masking observations in the NIRC 
narrowband filter set (with the exception of the $L$ band for W~Hya in 1999 Apr 25). 
Since calibration of these observations was not impacted by the blind tracking 
errors, we could obtain good contemporaneous angular diameters of our
targets at a range of discrete wavelengths across the grism passband.
Assuming that the tracking-induced instabilities in the optical 
transfer function were a smoothly-varying function with wavelength, we could
use our robust discrete UD angular diameters to pin the calibration at those wavelengths,
and perform a smooth interpolation in between to cover all passbands used 
in the dispersed experiment. In the case of W~Hya in 1999 Apr 25, we used
$L\,3.08$ 2-D making observations conducted at the same phase albeit one year earlier, on 1998 Apr 14 (cf. \citealt{WOO08}),
for one calibration of the GR60 grism.

Observations of the PSF reference stars were then used as check stars to verify the utility of this calibration process.
Although these check stars were typically smaller and fainter in the NIR than the targets, and therefore
yield proportionately higher relative errors on the diameter measurement, the 
results of applying our calibration methodology to these objects confirmed
that it was robust.
PSF reference stars had typical spectral types from late K to early M, and
(in comparison with Mira stars) these stars have compact atmospheres so that any diameter changes 
with wavelength are expected to be small.
After applying our new calibration strategy to these check stars, the results
verified these expectations and showed that the method produced 
a well-characterized constant diameter measurement over all three grism passbands.
This is shown in Fig. \ref{2_cen} for the reference star 2~Cen, observed contemporaneously with W~Hya in
1999 Apr 25. 

\subsection{Low resolution spectra}\label{spectra}

The observations of reference stars also allowed the spectral response of the optical system to be estimated.
We extracted the total flux as a function of wavelength for both science objects and reference stars,
making it possible to recover basic low resolution NIR spectra in the 1.0-4.0\,$\mu$m range (see Table \ref{grisms}) contemporaneously to our visibility measurements.
Since the spectral calibration suffers from the same observational difficulty which compromised
the calibration process of the visibilities outlined in Section \ref{observations}, it was not possible to obtain calibrated absolute fluxes from
our measurements.

\section{Results}\label{results}

Figures \ref{w_hya_2} to \ref{ocet_jul02} show the UD diameters of W~Hya, R~Leo and $o$~Cet as a function of wavelength (UD(${\lambda}$)),
together with contemporaneous spectra.
The gaps in the data correspond to regions of attenuated signal due to the telluric absorption bands,
with the exception of the gaps at wavelengths greater than $\sim$\,3.5\,$\mu$m, which are caused by saturation of the detector due to high flux from the sky background.
Also shown in the Figures are the best fitting UD($\lambda$) from dynamic model atmospheres with the respective model spectra
which are described and compared with the data in Section \ref{comp}.

The UDs show wavelength-dependent features that
are consistent with expected opacity changes due to molecular abundances.
Although some VO is responsible for UD($\lambda$) variations in the $J$-band,
the majority of the features are caused by the presence of H$_2$O (plus contributions of CO and OH) in the stellar atmosphere (see, e.g., \citealt{TLSW,WIT08}).
There is a very close anti-correlation between the spectral features and the apparent angular diameters at varying wavelengths.
We find local minima in the UD diameters at $\sim$\,1.3--1.4, 1.6--1.7, and 2.2\,$\mu$m, corresponding to the regions of increased flux in the spectra.
 This anti-correlation can be explained by a molecular blanketing model, where opaque layers above the continuum-forming layers (the photosphere)
 lead to a perceived UD angular diameter increase at certain wavelengths. This wavelength-dependent  absorption is also responsible for the 
 diminished flux in the same bandpasses. The less contaminated bandpasses, which are closer to sampling the continuum-forming layers,
 exhibit smaller measured UD diameters and higher fluxes.
 These results can be readily compared with the findings for S~Ori of \cite{WIT08}, which show UD diameter minima in the same wavelength
 regions and present a very close match regarding the overall form of the UD($\lambda$) curve. 
 
\begin{figure*}
\begin{center}
\epsscale{1}

\plotone{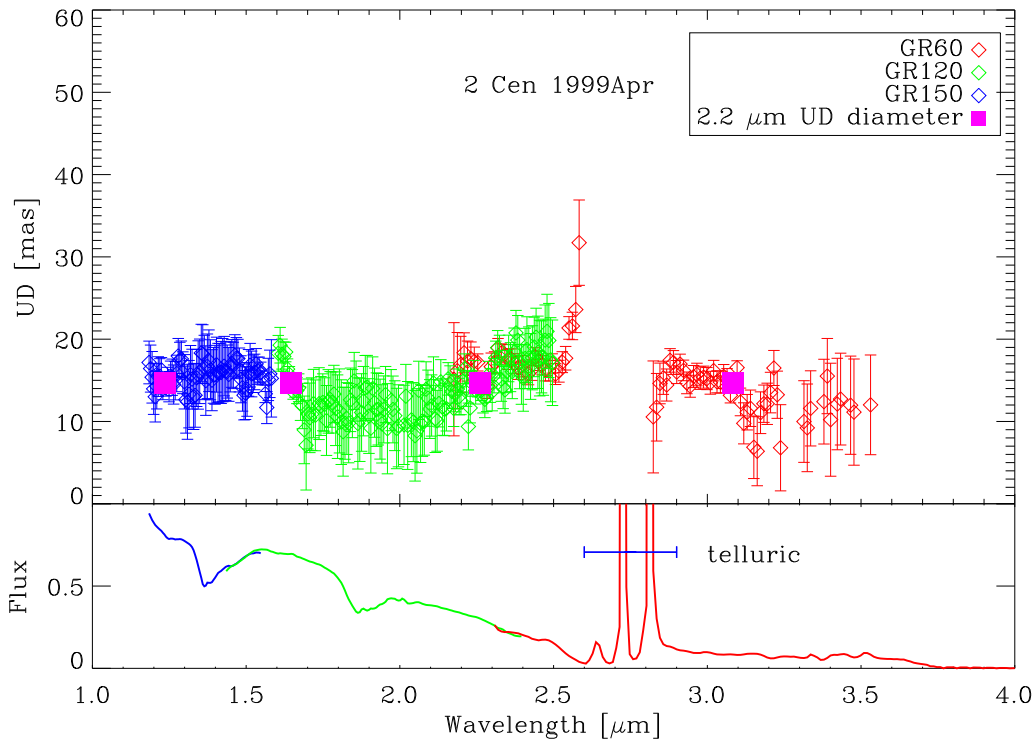}

\caption{{\it  Top}: UD angular diameters for the reference star 2~Cen as a function of wavelength observed on 1999 Apr 25
with the GR150, GR120 and GR60 grisms (blue, green and red diamonds respectevely). The magenta filled squares show the 2.2\,$\mu$m diameter of 14.7\,mas (cf. \citealt{DUM98}) which are used for calibration (see Section \ref{UDC}.)
{\it Bottom:} Low resolution spectra for 2~Cen observed contemporaneously with the GR150, GR120 and GR60 grisms (blue, green and red lines respectively).
The flux per unit wavelength is in arbitrary units and normalized at the $H$-band.
}
\label{2_cen}
\end{center}

\end{figure*}

For the partially-resolved reference star 2~Cen we assumed a UD diameter of 14.7\,mas (cf. \citealt{DUM98}) at the same wavelengths
used to pin the Mira grism data (cf. Figure \ref{2_cen}). Note that some variation of size with wavelength is expected for this M4.5 III star.
The corresponding spectrum is also fairly smooth, exhibiting only small H$_2$O absorption features at $\sim$\,1.4 and 1.9\,$\mu$m.

Figures \ref{w_hya_2} and \ref{rleo_1}  also show significant cycle-to-cycle and phase effects, which tend to be more pronounced towards shorter wavelengths.
This effect is predicted by the models, as discussed in the next Section.

\section{Comparison with models}\label{comp}

In this section the predictions of recent dynamic model atmospheres of M-type Miras are compared with the present observations.
The distances to the model stars were chosen such that the UD($\lambda$)$_{\rm model}$ overlaps our grism data in the wavelength region 2.2-2.5\,$\mu$m.
These distances do not guarantee good fitting to model photometry since
we only consider a comparison between the functional form of the model predictions as a function of wavelength and the observed data.
A comparison between model photometric and angular diameter fit distances will be given in a future paper describing the models.   
The new models, spanning several pulsation periods, are based on a new code for treating pulsation and
calculating atmospheric stratifications, as described below.

 The code of \cite{KEL06} was used for constructing two series of self-excited pulsation models with parameters
close to the well-studied prototypes $o$~Cet and R~Leo: (i) o54 model 
series with period $\sim$\,330\,days (non-pulsating ``parent star'' with mass 1.1\,M$_{\sun}$, 
luminosity 5400\,L$_{\sun}$, radius 216\,R$_{\sun}$, effective temperature 3370\,K, solar 
abundances with Z=0.02); (ii) r52 model series with period $\sim$\,307\,days 
(1.1\,M$_{\sun}$, 5200\,L$_{\sun}$, 209\,R$_{\sun}$, 3400\,K, solar Z=0.02). The temperature 
structure of the atmospheric layers was calculated with an opacity-sampling 
technique accounting for major molecular species blanketing M-type atmospheres 
as well as for dust opacity (\citealt{IRE08}, henceforth called ISW08). The o54 series covers three
time intervals, each containing 1.25 (interval hx, hereafter o54(hx)), 1.25 (jx) and 3.75 (fx) successive cycles in 
phase steps of about 0.1 times the pulsation period. For the r52 series, there 
are two time intervals with 2.25 (gx) and 2.5 (fx) successive cycles.

Instantaneous relaxation of hot matter behind shock fronts is assumed, and the 
approximation of local thermodynamical equilibrium was adopted except for 
scattering processes affecting continuous, TiO line and dust extinction (see
ISW08 for details). All model atmospheres are cut off at 5 parent star radii, 
which is considered to be the transition zone to a wind-dominated circumstellar 
shell. Though W~Hya has a longer period than these prototype model 
series, we would expect that scaling their atmospheric geometry structure to 
the larger size of a longer-period Mira provides a fair approximation for this 
star.

Inspection of the model series shows that, apart from the strong dependence of
the shape of UD($\lambda$)$_{\rm model}$ curves with optical phase, these curves also vary quite
noticeably between cycles. Whilst an observed small set of isolated
UD($\lambda$)$_{\rm obs}$ values or an observed narrow wavelength interval of the UD($\lambda$)$_{\rm obs}$
curve might readily be fitted by model predictions in different cycles, close
agreement is much harder to achieve over a broad wavelength range, as we find with our NIR observations. 
This reflects the varying structure of 
upper atmospheric layers, depending on the shock front position in the present 
cycle as well as on the shock front history of preceding cycles, which results 
in varying strengths of different molecular absorption features. Systematic
observations of monochromatic cycle-to-cycle size variations of Miras are not 
available, but the present observations of W~Hya in 1999Feb (phase 0.58) and 
2000Jan (phase 1.53) are almost one full cycle apart and clearly show 
noticeable differences of the UD($\lambda$)$_{\rm obs}$ curve, which are significantly larger 
than the small 0.05 phase-difference effect expected from models (Fig. \ref{w_hya_2}).

\subsection{W Hya}\label{whyacomp}

\begin{figure*}
\begin{center}
\epsscale{0.8}

\plotone{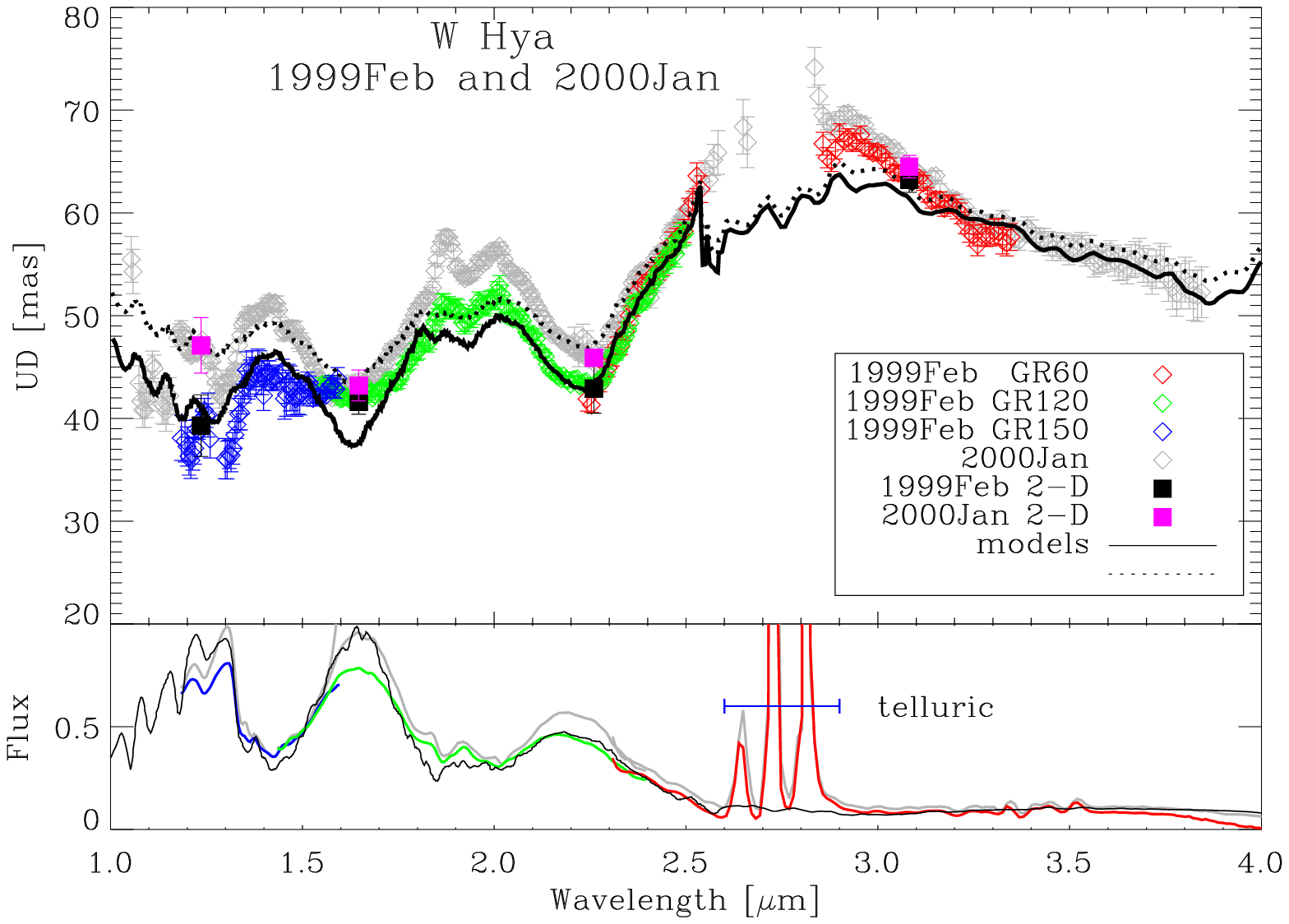}

\caption{{\it  Top}: UD angular diameters for W~Hya as a function of wavelength observed on 1999 Feb 5 ($\Phi=0.58$) and 2000 Jan 25 ($\Phi=1.53$)
with the GR150, GR120 and GR60 grisms (blue, green and red diamonds respectively for 1999Feb, and 
grey diamonds for Jan2000). The black solid and dashed lines show the fairly well fitting models of the o54(fx) series at phases
2.51 and 3.49. The black and magenta filled squares show the narrow-band two-dimensional UD angular diameters observed on the same nights which are used for calibration (see Section \ref{UDC}.)
{\it Bottom:} Low resolution spectra for W~Hya observed contemporaneously with the GR150, GR120 and GR60 grisms (blue, green and red lines respectively for 1999Feb, and grey lines for Jan2000).
The flux per unit wavelength is in arbitrary units and normalized at the $H$-band.
The black line shows the spectrum predicted by the model of the o54(fx) series at phase
2.51 (for the sake of clarity and since both model spectra differ only slightly, only one model spectrum is shown).
}
\label{w_hya_2}
\end{center}
\end{figure*}

\begin{figure*}[htbp]\begin{center}
\epsscale{0.8}

\plotone{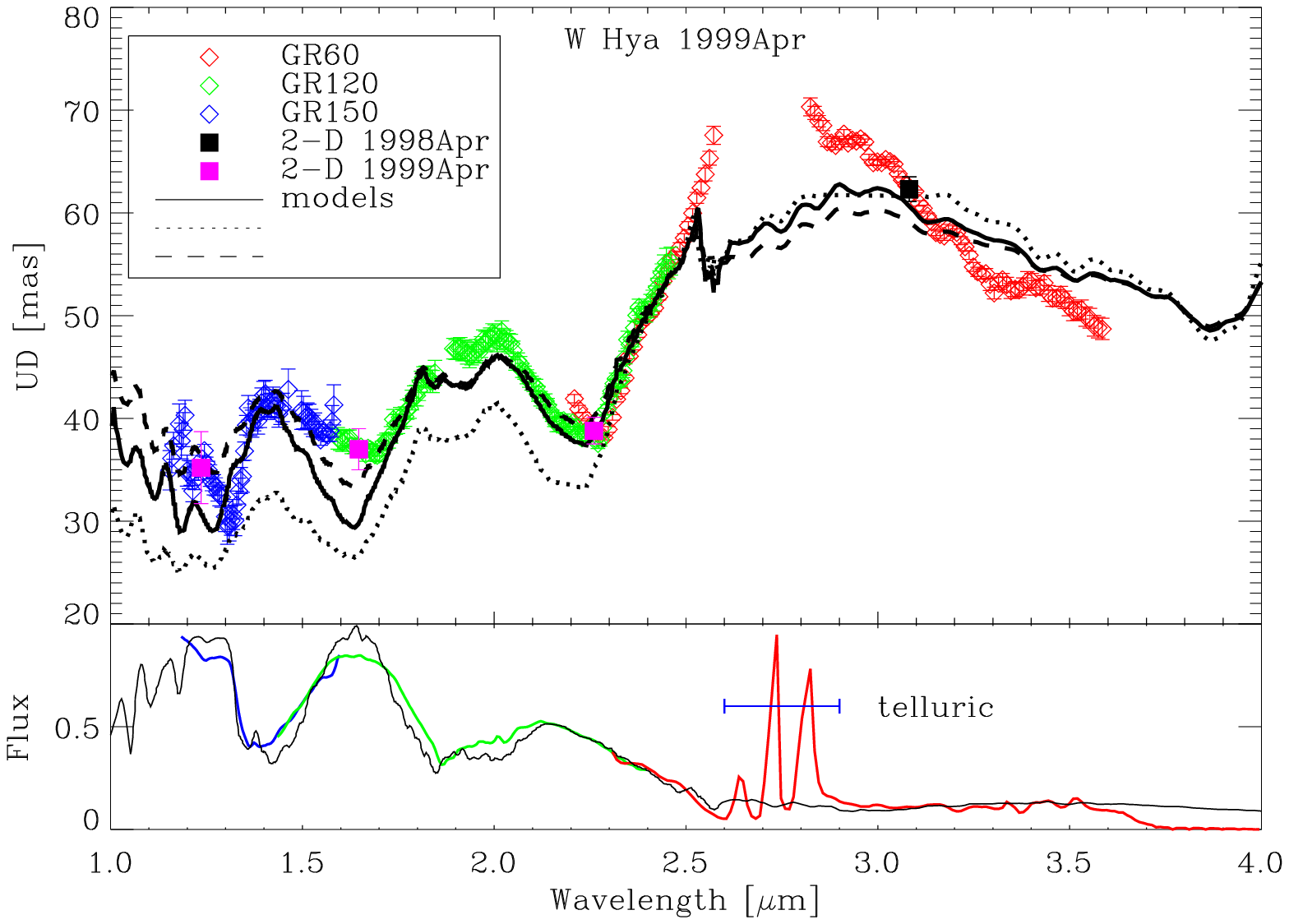}

\caption{Same as Fig. \ref{w_hya_2} but for W~Hya observed on 1999 Apr 25 ($\Phi=0.79$) only. Also shown is  the UD($\lambda$) curve as predicted
by the stellar models discussed in Section \ref{comp} as a black solid, dotted and dashed lines for the phases 2.70, 2.80 and 3.70 from the o54(fx) model series, 
respectively. The model spectrum corresponds to the phases 2.70 from the o54(fx) model series.
}
\label{w_hya_2a}
\end{center}
\end{figure*}

We compared the 3 phases of observation of W~Hya with predictions of different
cycles of both the o54 and r52 model series. Whilst no good agreement was found
with available r52 cycles, there is reasonable agreement with one 
modelled phase sequence of the o54 series when the phase assignment of the 
model series is estimated to be slightly earlier, say of the order of 0.05 to 0.1 
(which is well within the typical model vs. star phase uncertainty of at least
0.1). Even better agreement is achieved between our phase 0.58 observation and models with nearby phases 4.55 
and 4.61 but, since subsequent model cycles are presently not available, the
observed phases 0.79 and 1.53 cannot be compared to models in this case.  

For good agreement with observed UD diameters, the model star had to be 
put at a ``fit-distance'' of about 78 pc, as compared to $104\pm10$\,pc for the 
real star (\citealt{WHI08}), where the difference reflects the fact that 
W Hya has a noticeably longer period (385\,days) and hence is larger than the 
o54 model star (330\,days, parent star radius 216\,R$_{\sun}$).

\begin{figure*}
\begin{center}
\epsscale{0.8}

\plotone{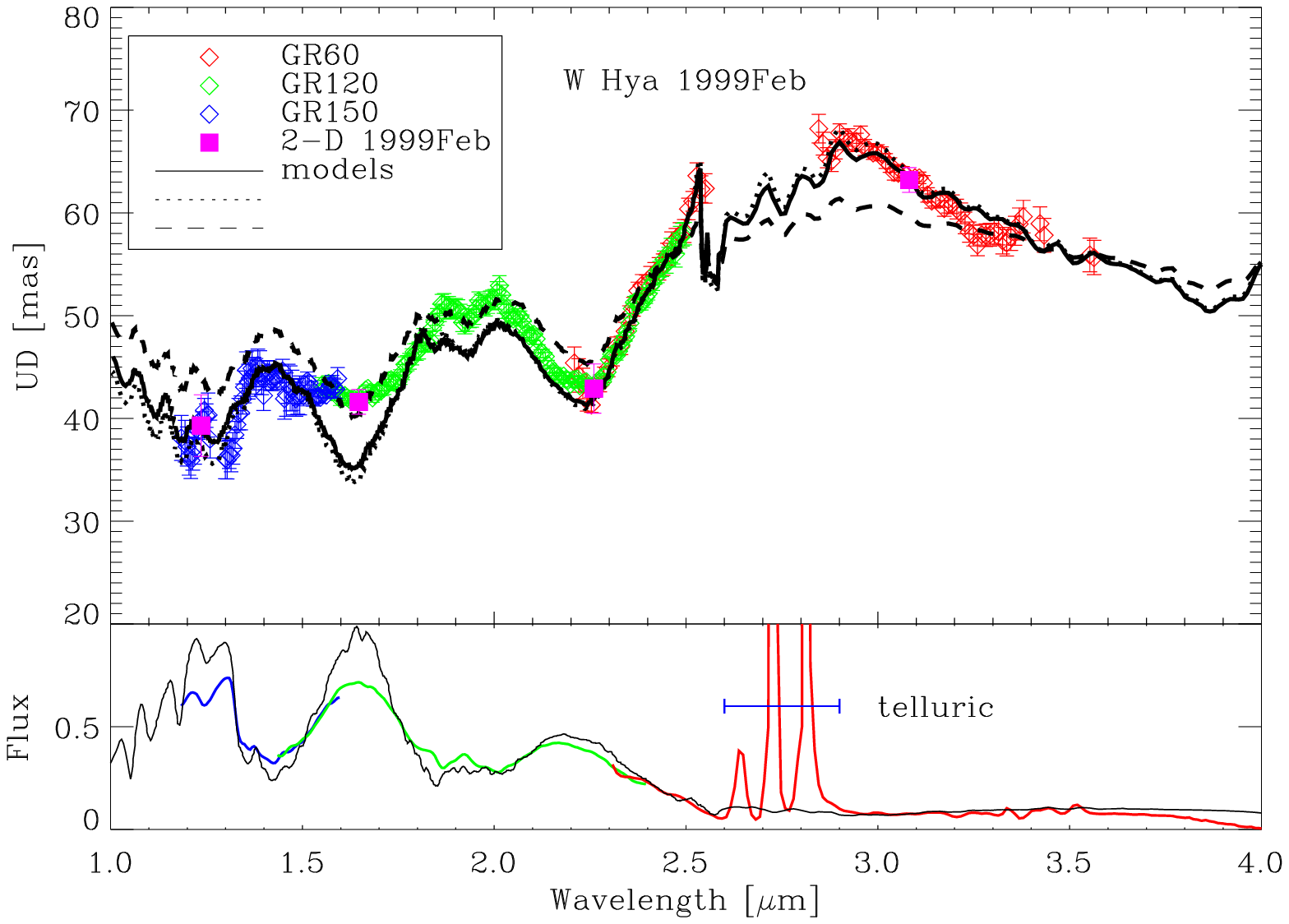}

\caption{Same as Fig. \ref{w_hya_2} but for W~Hya observed on 1999 Feb 5 ($\Phi=0.58$) only. Also shown are the UD($\lambda$) curves as predicted
by the stellar models discussed in Section \ref{comp} as black solid, dotted and dashed lines for the phases 4.55, 4.61 from the o54(fx) model series as well
as the model at 3.59 of the preceding cycle for demonstrating significant cycle-to-cycle effects, respectively.
The model spectrum corresponds to the phases 4.55 from the o54(fx) model series.
}
\label{w_hya_3}
\end{center}
\end{figure*}

In Figure \ref{w_hya_2} we show the 1999Feb and 2000Jan ($\Phi=0.58$ and $\Phi=1.53$, respectively) observation together with UD($\lambda$) curves of o54(fx) models with phases 2.51 and 3.49 of successive cycles.
Figure  \ref{w_hya_2a} shows the 1999Apr observation ($\Phi=0.79$) together with the phase 2.70 model, as well as the phase 2.80 and 3.70 models for demonstrating significant phase and cycle-to-cycle effects.
Figure  \ref{w_hya_3} shows the 1999Feb observation of W Hya ($\Phi=0.58$) with phase 4.55 and 4.61 models plus the phase 3.59 model of the preceding cycle which gives a less satisfactory overall fit. 
We note that the cycle containing phases 3.49 to 3.70 has generally smaller UD values than the other cycles used in 
Figs. \ref{w_hya_2} to \ref{w_hya_3}, i.e. the model star appears smaller in this cycle and had to be scaled accordingly.

\subsection{R Leo}\label{rleocomp}

\begin{figure*}
\begin{center}
\epsscale{0.8}

\plotone{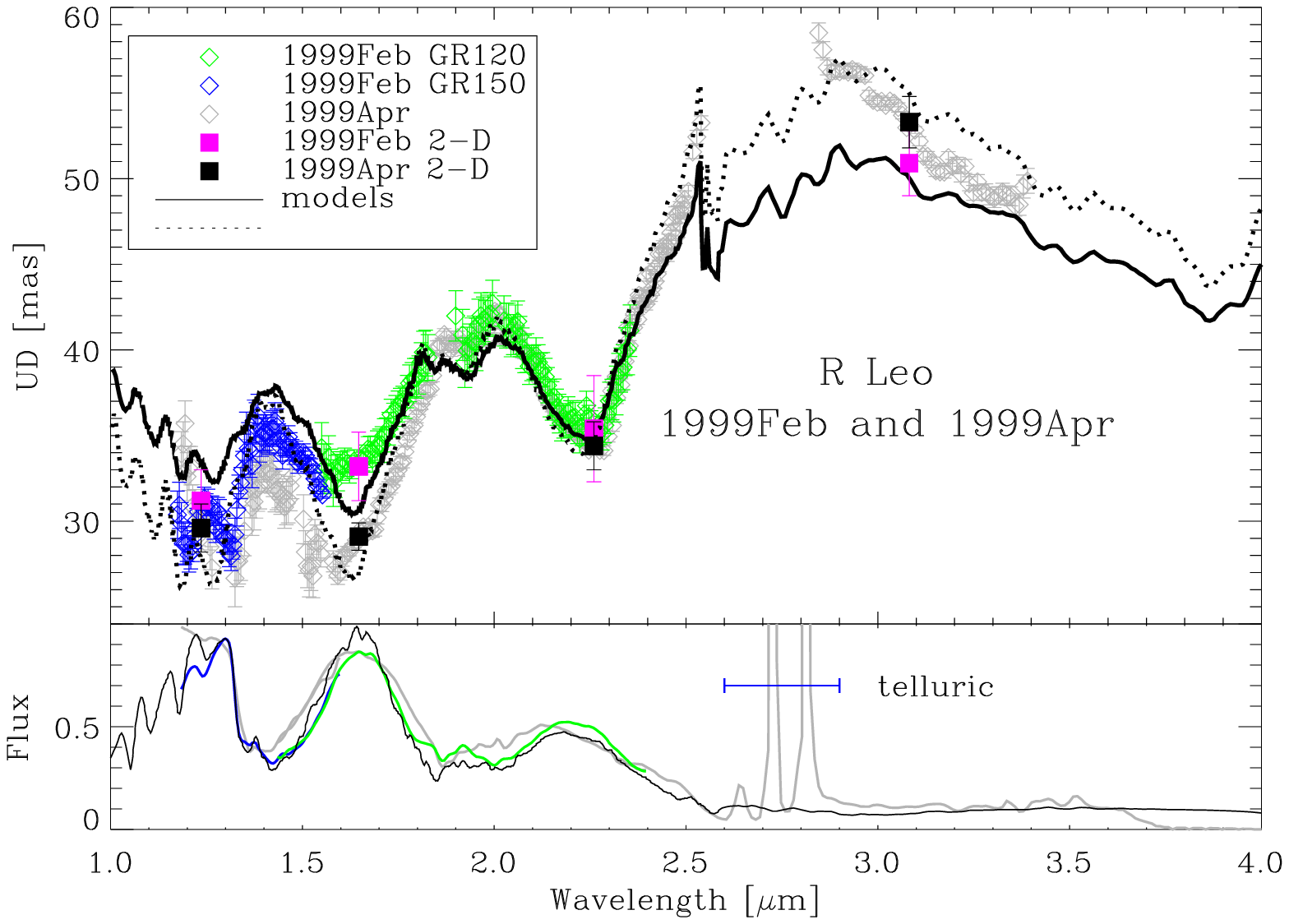}

\caption{Same as Fig. \ref{w_hya_2} but for R~Leo observed on 1999 Feb 04 ($\Phi=0.49$) and on 1999 Apr 25 ($\Phi=0.75$) with the GR150 and GR120 grisms (blue and green symbols respectively for 1999Feb, and grey symbols for 1999Apr).
The model UD($\lambda$) curves shown as a solid and a dashed black line correspond to the phases 2.51 and 2.70 of the o54(fx) model series.
The model spectrum corresponds to the phases 2.70 from the o54(fx) model series.}
\label{rleo_1}
\end{center}

\end{figure*}

For the observations of R~Leo at two phases, none of the four available cycles of 
the r52 model series (which was originally designed to describe R Leo) matches 
the observed UD($\lambda$) curve well, whereas two cycles of the o54 series with a 
slightly longer period give a fair fit. For good agreement with observations, the model star was set at a distance of $\sim93$\,pc,
to be compared with parallaxes 
given by \citet[$120\pm14$\,pc]{GAT92} and \citet[HIPPARCOS 
$73\pm6$\,pc]{LEE07} and with a weighted-mean distance suggested by \citet[$111\pm17$\,pc]{WHI08}. 
Here, similar to the case of W~Hya, we note that 
the $\Phi=0.75$ observation is in the phase range of pronounced UD($\lambda$) phase dependence when the Mira light curve increases very steeply.
Fig. \ref{rleo_1} shows the observed UD($\lambda$) curves together with model curves.
These give a fair fit within the $HKL$ bands, whereas the modeled UD diameters in the $J$ bandpass is about 15 percent larger than the observed value at phase 0.49.

\subsection{$o$ Cet}\label{ocetcomp}

\begin{figure*}
\begin{center}
\epsscale{0.8}

\plotone{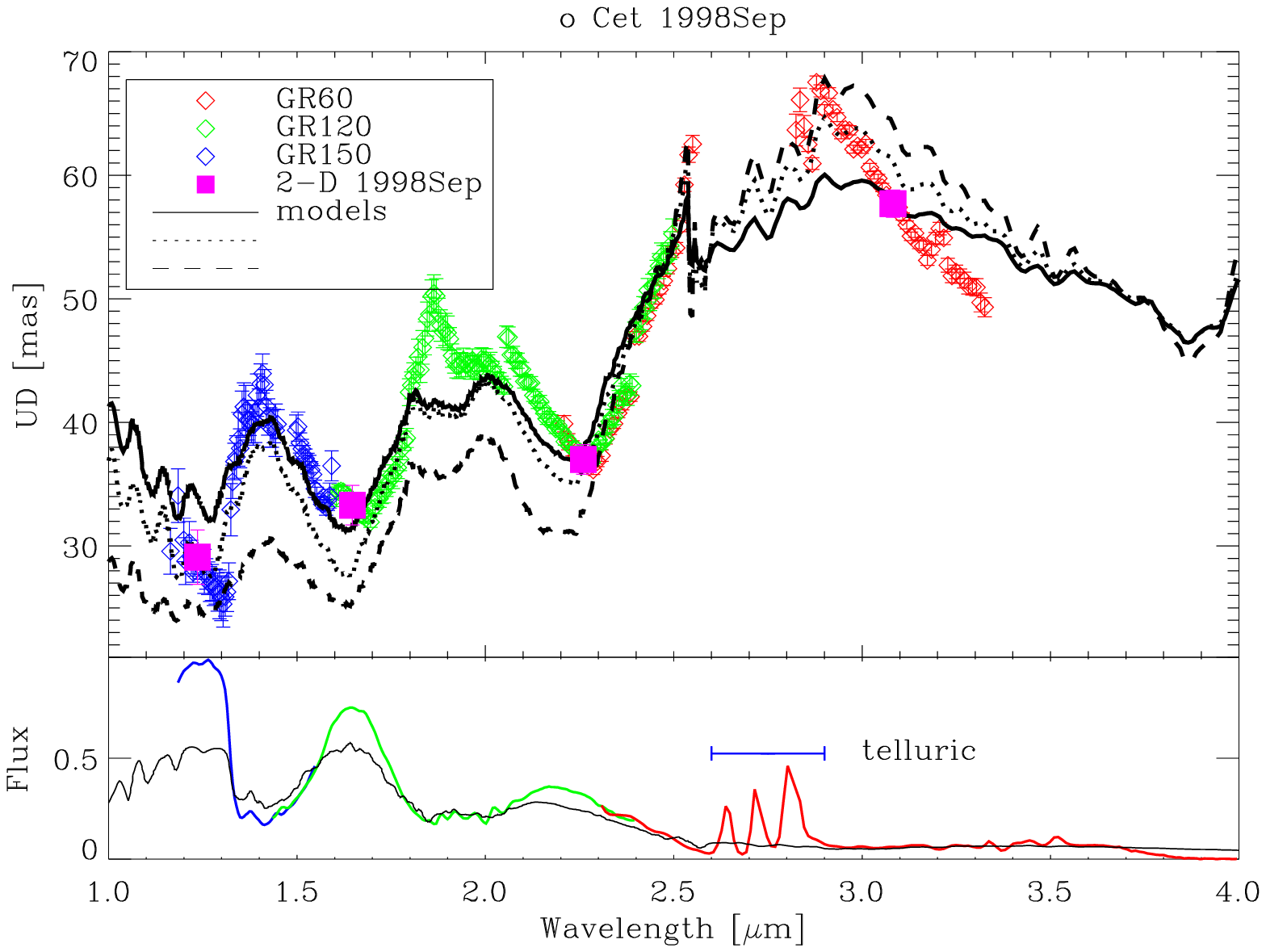}

\caption{Same as Fig. \ref{w_hya_2} but for $o$~Cet observed on 1998 Sep 29 ($\Phi=0.71$). The model UD($\lambda$) curves shown
as solid, dotted, and dashed black lines correspond to the phases 0.70, 1.70 and 1.80 from the o54(hx) model series, respectively.
The model spectrum corresponds to the phases 0.70 from the o54(fx) model series.}
\label{ocet_sep98}
\end{center}

\end{figure*}

\begin{figure*}
\begin{center}
\epsscale{0.8}

\plotone{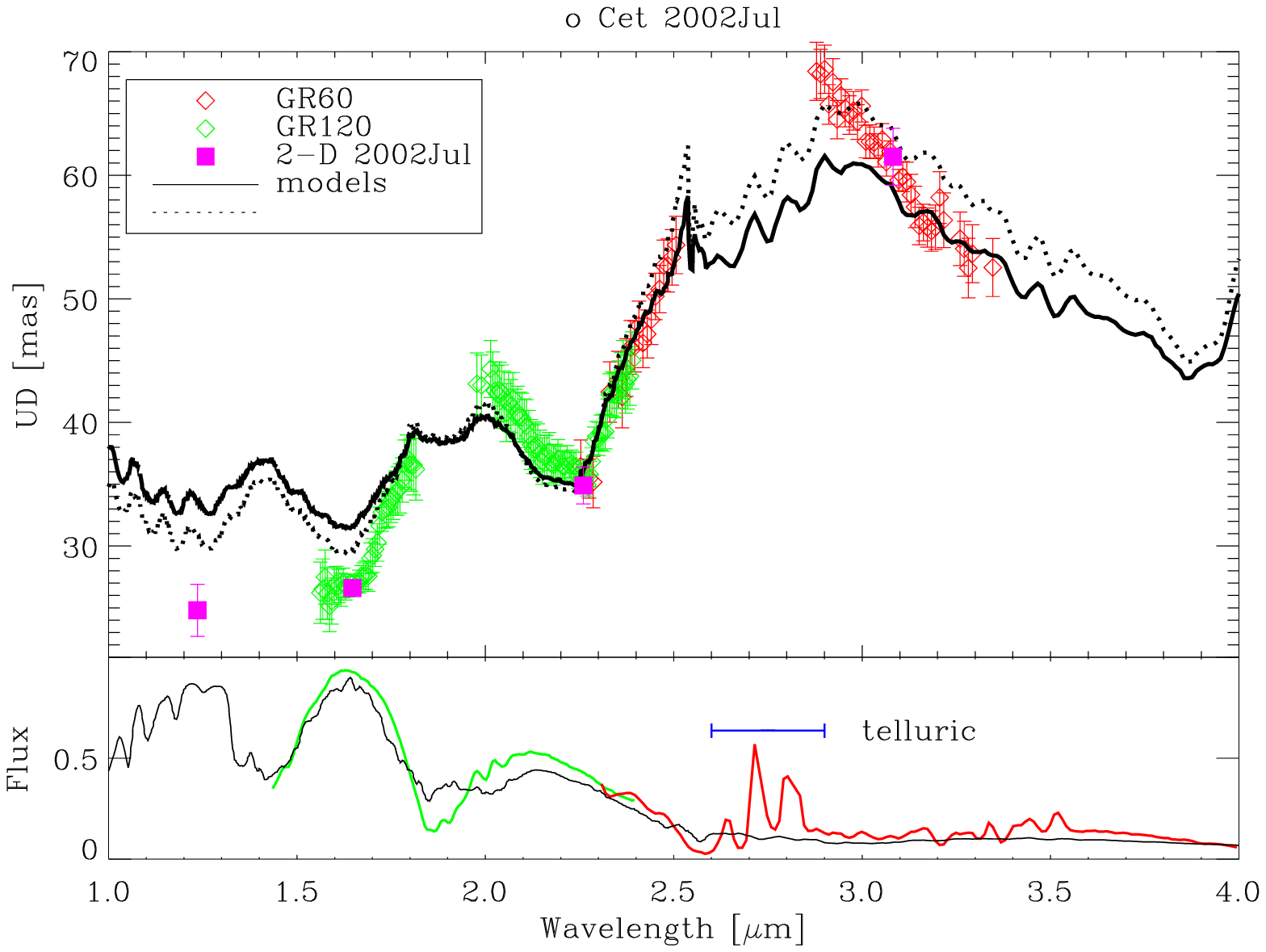}

\caption{Same as Fig. \ref{w_hya_2} but for $o$~Cet observed on 2002 Jul 23 ($\Phi=4.98$). The model UD($\lambda$) curves shown
as solid, and dotted black lines correspond to the identical phases 5.00 and 5.99 from two successive cycles of the the r52(fx) model series, respectively.
The model spectrum corresponds to the phases 5.00 from the o54(fx) model series.}
\label{ocet_jul02}
\end{center}

\end{figure*}

For $o$~Cet, observations were made at two phases in two quite distinct cycles
(1998Sep and 2002Jul) so that the phase-cycle-effects cannot be studied readily in this
case. Two basic problems occur when we fitted available models to the observed UD($\lambda$) curves. 
(i) The model parent stars have to be placed at a distance of $\sim$80\,pc (o54) or $\sim$70\,pc (r52), both of which are
marginally smaller than values given in the literature, e.g. 92 $\pm10$\,pc after \cite{WHI08}. 
(ii) The decrease of the UD diameter measured  on 1998Sep at the 
long-wavelength side of the strong 3\,$\mu$m water band tends to be noticeably steeper than predicted by any models.
Figure \ref{ocet_sep98} shows the 1998Sep observation of $o$~Cet at $\Phi=0.71$ together with same phase models of two successive cycles,
showing cycle effects, and a model with 0.1 later phase, demonstrating the strong phase effects around $\Phi=0.8$. In Fig. \ref{ocet_jul02}
the ($\Phi=4.98$) observation of $o$ Cet is shown with two model curves of the r52(fx) series.

\section{Summary}\label{sum}

We have measured the wavelength dependent diameters of three
Mira variables stars from 1.0-3.4\,$\mu$m  and their respective spectra at up to 3 different epochs, the first study of this type in
the near-infrared. All observed stars show strong variations of their UD angular diameters as a function of wavelength,
often exhibiting a factor of $\sim$\,2 in UD diameter between 1.0\,$\mu$m and 3.0\,$\mu$m.
We find that the UD($\lambda$) relation shows variations with phase and the pulsation cycle,
 revealing the complexity of cycle and phase-dependent opacity contributions from molecules, predominantly H$_2$O and CO, in different layers. \\

We found that there is fair agreement between the measured UD($\lambda$) relationship and spectra, and those
predicted by theoretical models designed to represent o Cet (see ISW08).
When comparing the two stars whose parameters are thought to resemble the model input parameters the most,
R~Leo and $o$~Cet, with model predictions (\citealt{IRE08}), we find that the model UD diameters are slightly too small
throughout the $1.1-3.8\,\mu$m region.

Our data also shows good agreement with results for the Mira star S Ori by \cite{WIT08},
in that the minimum UD angular diameters are smallest at 1.3-1.4 and 1.6-1.7\,$\mu$m
and increase by a factor of 1.4-1.5 around 2.0\,$\mu$m.
For a full comparison with published interferometric UD angular diameters in continuum bandpasses, see \cite{WOO08}. 

Further work with these data, including asymmetry studies, will be presented
in subsequent publications.

\acknowledgments

This work has been supported by grants from the National Science Foundation,
 the Australian Research Council and the Deutsche
Forschungsgemeinschaft (MS). The data presented herein were
obtained at the W.M. Keck Observatory, which is operated as a
scientific partnership among the California Institute of Technology,
the University of California and the National Aeronautics and Space
Administration.  The Observatory was made possible by the generous
financial support of the W.M. Keck Foundation.
We acknowledge with thanks the variable star observations from the AAVSO International Database contributed by observers worldwide and used in this research.
We also thank Albert Jones and Peter Williams for the W Hya light curve data.

\bibliographystyle{apj}
\bibliography{REFPAPER}

\end{document}